\documentclass[energies,article,accept,moreauthors,pdftex,12pt,a4paper]{mdpi}
\lastpage{x}
\doinum{10.3390/------}
\pubvolume{6}
\pubyear{2013}
\history{Received: 1 July 2013; in revised form: 17 August 2013 / Accepted: 21 August 2013 / \linebreak Published: xx}

\usepackage[utf8]{inputenc}
\usepackage{import}
\usepackage{xspace}
\usepackage{amssymb}
\usepackage{amstext}
\usepackage{amsmath}
\usepackage{amsthm}
\usepackage{graphicx}
\usepackage{color}
\usepackage[table]{xcolor}
\usepackage{comment}
\usepackage{soul}
\usepackage{booktabs}
\usepackage{pdflscape}

\newcommand{\ann}{ANN\xspace}
\newcommand{\anns}{ANNs\xspace}
\newcommand{\openhab}{openHAB\xspace}
\newcommand{\caes}{CAES\xspace}
\newcommand{\mcs}{MCS\xspace}
\newcommand{\knx}{KNX\xspace}

\newcommand{\smlhouse}{SMLhouse\xspace}
\newcommand{\smlsystem}{SMLsystem\xspace}

\newcommand{\best}{BEST\xspace}
\newcommand{\comb}{COMB-EXP\xspace}
\newcommand{\combeq}{COMB-EQ\xspace}
\newcommand{\model}[1]{\theta_{#1}}
\newcommand{\data}{\mathcal{D}}
\newcommand{\combw}{\alpha}

\newcommand{\sequence}[1]{\bar{#1}}
\newcommand{\mean}[1]{\mathbb{E}[#1]}

\Title{Towards Energy Efficiency: Forecasting Indoor Temperature via Multivariate Analysis}

\Author{Francisco Zamora-Mart\'inez *, Pablo Romeu, Paloma Botella-Rocamora and Juan Pardo}
\address[1]{Escuela Superior de Ense\~nanzas T\'ecnicas, Universidad CEU Cardenal Herrera, C/ San Bartolom\'e 55, Alfara del Patriarca  46115, Valencia, Spain; E-Mails: pablo.romeu@uch.ceu.es (P.R.); pbotella@uch.ceu.es (P.B.-R.); juan.pardo@uch.ceu.es (J.P.)}

\corres{E-Mail: francisco.zamora@uch.ceu.es; \linebreak Tel.: +34-961-36-90-00 (ext. 2361).}

\abstract{The small medium large system (\smlsystem) is a house built at the Universidad CEU Cardenal Herrera (CEU-UCH) for participation in the Solar Decathlon 2013 competition. Several technologies have been integrated to reduce power
 consumption. One of these is a forecasting system based on artificial neural
 networks (\anns), which is able to predict indoor temperature in the near
 future using captured data by a complex monitoring system as the input. A study
 of the impact on forecasting performance of different covariate combinations
 is presented in this paper. Additionally, a comparison of ANNs with the standard
 statistical forecasting methods is shown. The research in this paper has been
 focused on forecasting the indoor temperature of a house, as it is directly
 related to HVAC---heating, ventilation and air conditioning---system
 consumption. HVAC systems at the \smlsystem house represent $53.89\%$ of the
 overall power consumption. The energy used to maintain temperature was measured to
 be $30\%$--$38.9\%$ of the energy needed to lower it. Hence, these forecasting
 measures allow the house to adapt itself to future temperature conditions by
 using home automation in an energy-efficient manner. Experimental results
 show a high forecasting accuracy and therefore, they might be used to
 efficiently control an HVAC system.}


\begin{document}

\section{Introduction}

\label{sec:introduction}
Nowadays, as the Spanish Institute for Diversification and Saving of Energy
(IDAE) \cite{IDAE} of the Spanish Government says, energy is becoming a precious
asset of incalculable value, which converted from electricity, heat or fuel,
makes the everyday life of people easier and more comfortable. Moreover, it
is also a key factor to make the progress of industry and
business feasible.

Spanish households consume $30\%$ of the total energy expenditure
of the country \cite{IDAE}. In the European Union (EU), primary energy consumption
in buildings represents about $40\%$ of the total \cite{Ferreira}. In the whole
world, recent studies say that energy in buildings also represents a $40\%$ rate
of the total consumed energy, where more than half is used by heating, ventilation and air conditioning (HVAC) systems \cite{Alvarez}.

Energy is a scarce resource in nature, which has an important cost, is
finite and must be shared. \linebreak Hence, there is a need to design and implement new
systems at home, which should be able to produce and use energy efficiently and
wisely, reaching a balance between consumption and streamlined comfort. A person
could realize his activities much easier if his comfort is ensured and there are
no negative factors (e.g., cold, heat, low light, noise, low air quality, {\it etc.})
to disturb him. With the evolution of technology, new parameters have become more
controllable, and the requirements for people's comfort level \linebreak have increased.

Systems that let us monitor and control such aspects make it necessary to refer
to what in reference \cite{Arroyo} is called ``Ambient Intelligence'' (AmI). This refers
to the set of user-centered applications that integrate ubiquitous and
transparent technology to implement intelligent environments with natural
interaction. The result is a system that shows an active behavior (intelligent),
anticipating possible solutions adapted to the context in which such a system is
located. The term, home automation, can be defined as it is mentioned
in reference \cite{Sierra}, as the set of services provided by integrated technology
systems to meet the basic needs of security, communication, energy management
and comfort of a person and his immediate environment. Thus, home automation can
be understood as the discipline which studies the development of intelligent
infrastructures and information technologies in buildings. In this paper, the concept of smart
 buildings is used in this way, as constructions that involve this kind
of solution.

In this sense, the School of Technical Sciences at the University CEU-UCH has built a solar-powered house, known as the Small Medium Large System (\smlsystem), which
integrates a whole range of different technologies to improve energy efficiency, allowing it to be a near-zero energy house. \linebreak The house has been constructed to participate in the 2012 Solar
Decathlon Europe competition. \linebreak Solar Decathlon Europe \cite{solar} is an
international competition among universities, which promotes research in the
development of  energy-efficient houses. The objective of the participating teams is to
design and build houses that consume as few natural resources as possible and
produce minimum waste products during their lifecycle. Special emphasis is
placed on reducing energy consumption and on obtaining all the needed energy
from the sun. The \smlsystem house includes a Computer-Aided Energy Saving
System (\caes). The \caes is the system that has been developed for the contest,
which aims to improve energy efficiency using home automation devices. This system
has different intelligent modules in order to make predictions about energy
consumption and production.

To implement such intelligent systems, forecasting techniques in the area of
artificial intelligence can be applied. Soft computing is widely used in
real-life applications~\cite{2009:Wu,2012:Taormina}. In fact, artificial neural networks
(\anns) have been widely used for a range of applications in the area of energy
systems \linebreak modeling \cite{Karatasou, 2006:energyandbuildings:ruano, Ferreira,
 2012:kdir:zamora}. The literature demonstrates their capabilities to work with
time series or regression, over other conventional methods, on non-linear
process modeling, such as energy consumption in buildings. Of special interest to
this area is the use of \anns for forecasting the room air temperature as a
function of forecasted weather parameters (mainly solar radiation and air
temperature) and the actuator (heating, ventilating, cooling) state or manipulated
variables, and the subsequent use of these mid-/long-range prediction models for
a more efficient temperature control, both in terms of regulation and energy
consumption, as can be read in reference \cite{2006:energyandbuildings:ruano}.

Depending on the type of building, location and other factors, HVAC systems
may represent up to $40\%$ of the total energy
consumption of a building \cite{Ferreira,Alvarez}. 
The activation/deactivation of such systems depends
on the comfort parameters that have been established, one of the most being indoor temperature, directly related to the notion of
comfort. Several authors have been working on this idea; \linebreak in reference \cite{Ferreira},
an excellent state-of-the-art system can be found. This is why the development of an \ann
to predict such values could help to improve overall energy consumption,
balanced with the minimum affordable comfort of a home, in the case that these values are
well anticipated in order to define efficient energy control actions.

This paper is focused on the development of an \ann module to predict the behavior
of indoor temperature, in order to use its prediction to reduce energy
consumption values of an HVAC system. The architecture of the
overall system and the variables being monitored and controlled are presented.
Next, how to tackle the problem of time series forecasting for the indoor temperature is
depicted. \linebreak Finally, the \ann experimental results are presented and compared to
standard statistical techniques.
Indoor temperature forecasting is an interesting problem which has been widely
studied in the literature, for example,
in~\cite{2008:EandB:Neto,Ferreira,Alvarez,2012:EandB:olderwurtel,2013:ESA:Mateo}. We
focus this work in multivariate forecasting using different
weather indicators as input features. In addition, two combinations of forecast models have been
compared.

In the conclusion, it is studied how the predicted results are integrated with the energy consumption parameters and comfort levels of the \smlsystem.

\section{\smlhouse and \smlsystem Environment Setup}

The Small Medium Large House (\smlhouse) and \smlsystem solar houses (more info about both projects can be found here: http://sdeurope.uch.ceu.es/)
have been built to participate in the Solar Decathlon 2010 and 2012 \cite{solar}, respectively, and aim to serve as prototypes for improving energy efficiency.
The competition focus on reproducing the normal behavior of the inhabitants of a house, requiring competitors to maintain comfortable conditions inside the house---to maintain temperature, CO$_2$ and humidity within a range, performing common tasks like using the oven cooking, watching television (TV), shower, {\it etc.}, while using as little electrical power as possible.

As stated in reference
\cite{DBLP:conf/infocom/PanYWXPPW12}, due to thermal inertia, it is more efficient to maintain a temperature of a room or building than cooling/heating it. Therefore, predicting indoor temperature in
the \smlsystem could reduce HVAC system consumption
using future values of temperature, and then deciding whether to activate the
heat pump or not to maintain the current temperature, regardless of its present value. To build an indoor temperature prediction module, a minimum of several weeks of sensing data are needed. Hence, the prediction module was trained using historical sensing data from the \smlhouse, 2010, in order to be applied in the \smlsystem.

The \smlhouse monitoring database is large enough to estimate forecasting models, therefore its database has been used to tune and analyze forecasting methods for
indoor temperature, and to show how they could be improved using different sensing data as
covariates for the models. This training data was used for the \smlsystem prediction
module.

The \smlsystem is a modular house built basically using wood. It was designed to
be an energy self-sufficient house, using passive strategies and water heating
systems to reduce the amount of electrical power needed to operate the house.

The energy supply of the \smlsystem is divided into solar power generation
and a domestic hot water (DHW) system. The photovoltaic solar system is responsible
for generating electric power by using twenty-one solar panels.
These panels are installed on the roof and at the east and west facades.
\linebreak The energy generated by this system is managed by a device to inject energy into the house,
or in case there is an excess of power, to the grid or a battery system.
The thermal power generation is performed using a solar panel that produces
DHW for electric energy savings.

The energy demand of the \smlsystem house is divided into three main groups: HVAC,
house appliances and lighting and home electronics (HE). The HVAC
system consists of a heat pump, \linebreak which is capable of heating or cooling water, in addition to a
rejector fan. Water pipes are installed inside the house, and a fan coil system distributes
the heat/cold using ventilation.
As shown in reference \cite{eebuildings}, \linebreak the HVAC system is the main contributor to
residential energy consumption, using $43\%$ of total power in U.S. households or $70\%$ of total power in European residential buildings.
In the \smlsystem,  \linebreak the HVAC had a peak consumption of up to $3.6$ kW
when the heat pump was activated and, as shown in Table~\ref{tab:consumption},
it was the highest power consumption element of the \smlsystem in the contest
with $53.89\%$ of total consumption. This is consistent with data from studies mentioned
 as the competition was held in Madrid (Spain) at the end of September.
The house has several energy-efficient appliances that are
used during the competition. Among them, there is a washing machine,
refrigerator with freezer, an induction hob/vitroceramic and a conventional oven.
Regarding the consumption of the washing machine and dishwasher,
they can reduce the \smlsystem energy demand due to the DHW system.
\linebreak The DHW system is capable of heating water to high temperatures. Then, when water
enters into these appliances, the resistor must be activated for a short time only to reach
the desired temperature. \linebreak The last energy-demanding group consists of several electrical outlets
(e.g., TV, computer, Internet router and others).

\begin{table}[H] \footnotesize \centering
 \begin{tabular}{cccc}
 \toprule
	{\bf System} & {\bf Power peak (kW)} & {\bf Total power (Wh)} & {\bf Percentage} \\
	 \midrule
 HVAC & $3.544$ & $37987.92$ & $53.89\%$ \\ 

 Home appliances & - & $24749.10$	& $35.11\%$ \\

 Lighting \& HE & $0.300$ & $7755.83$ & $11.00\%$ \\
 \bottomrule 
 \end{tabular}
 \caption{Energy consumption per subsystem. HVAC: heating, ventilation and air conditioning; HE: home electronics.\label{tab:consumption}}
\end{table}


Although the energy consumption of the house could be improved, the installed
systems let the \smlsystem house be a near-zero energy building, producing
almost all the energy at the time the inhabitants need it. This performance won the second place at the energy balance contest of the Solar Decathlon
competition. The classification of the Energy Balance contest can be found
 here: http://monitoring.sdeurope.org/index.php?action=scoring\&scoring=S4~.

\begin{figure}[H]
 \centering
 \includegraphics[width=0.8\textwidth]{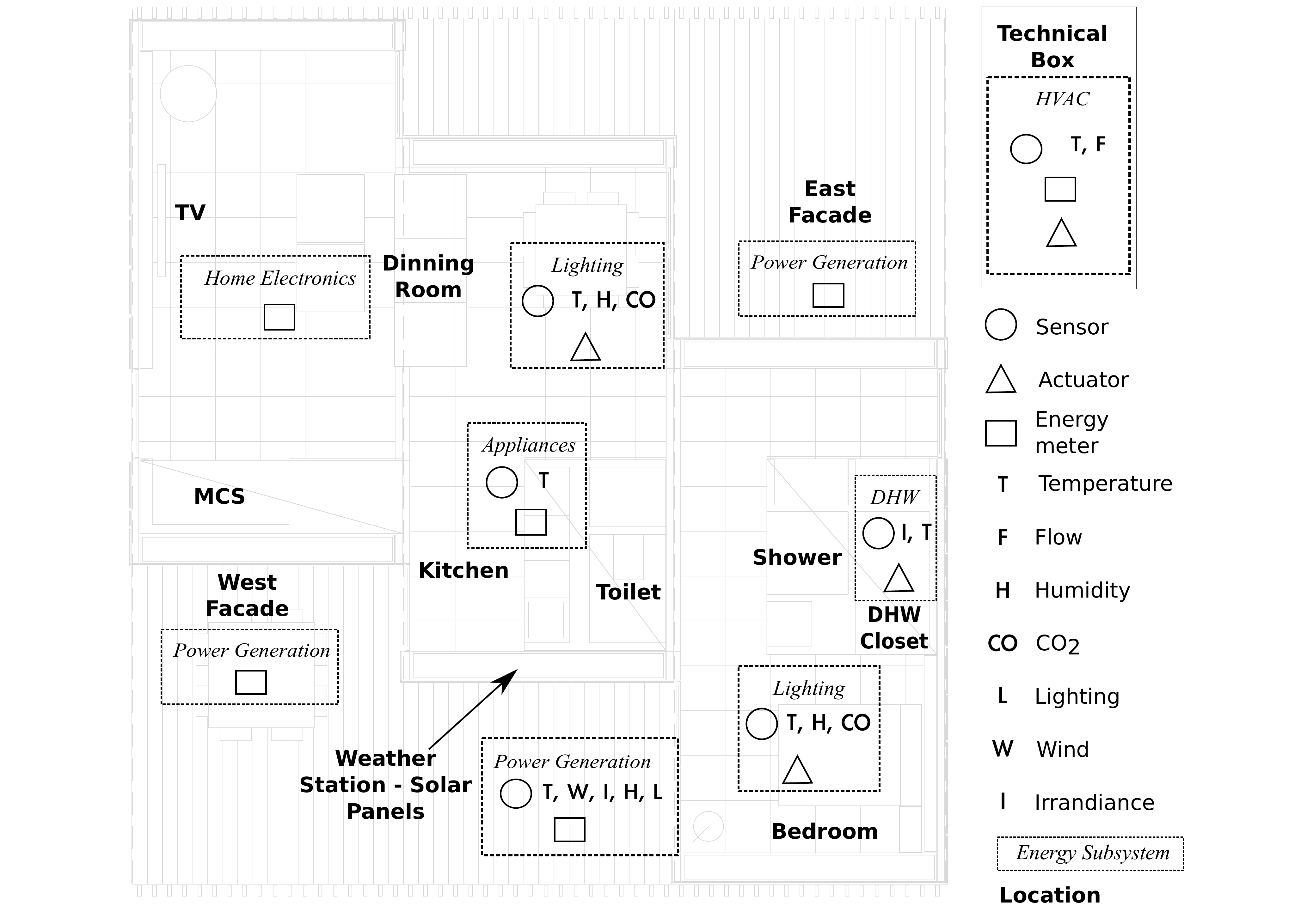}
 \caption{\smlsystem sensors and actuators map.\label{fig:plano_sensores}}
\end{figure}

A sensor and control framework shown in Figure~\ref{fig:plano_sensores}
has been used in the \smlsystem. It is operated by a Master Control
Server (\mcs) and the European home automation standard protocol
known as Konnex (\knx) (neither KNX nor Konnex are acronyms: http://ask.aboutknx.com/questions/430/abbreviation-knx)
 has been chosen for monitoring and sensing.
\knx modules are grouped by functionality: analog or binary inputs/outputs,
gateways between transmission media, weather stations, CO$_2$ detectors, {\it etc}.
The whole system provides $88$ sensor values and $49$ actuators.
In the proposed system, the immediate execution actions had been programmed to
operate without the involvement of the \mcs, such as controlling ventilation,
the HVAC system and the DHW system. Beyond this basic level, the \mcs
can read the status of sensors and actuators at any time and can
perform actions on them via an Ethernet gateway.

A monitoring and control software was developed following a three-layered scheme.
In the first layer, data is acquired from the \knx bus using a \knx-IP (Internet Protocol) 
 bridge device.
The Open Home Automation Bus  (\openhab)~\cite{openhab} software performs
the communication between \knx and our software.
In the second layer, it is possible to find a data persistence module that
has been developed to collect the values offered by \openhab with a
sampling period of 60 s. Finally, the third layer is
composed of different software applications that are able to
intercommunicate: a mobile application has been developed to let the
user watch and control the current state of domotic devices;
and different intelligence modules are being developed also, for instance,
the \ann-based indoor temperature forecasting module.

The energy power generation systems described previously are monitored by a software
controller. It includes multiple measurement sensors, including the voltage and
current measurements of photovoltaic panels and batteries. Furthermore, the
current, voltage and power of the grid is available. The system power
consumption of the house has sensors for measuring power energy values for each
group element. The climate system has power consumption sensors for the whole
system, and specifically for the heat pump. The HVAC system is composed of
several actuators and sensors used for operation. Among them are the inlet and
outlet temperatures of the heat rejector and the inlet and outlet temperatures
of the HVAC water in the \smlsystem. In addition, there are fourteen switches for
internal function valves, for the fan coil system, for the heat pump
and the heat rejector. The DHW system uses a valve and a pump to control water
temperature. Some appliances have temperature sensors which are also monitored.
The lighting system has sixteen binary actuators that can be operated manually by
using the wall-mounted switches or by the \mcs. The \smlsystem has indoor
sensors for temperature, humidity and CO$_2$. Outdoor sensors are also
available for lighting measurements, wind speed, rain, irradiance and
temperature.

\section{Time Series Forecasting}\label{sec:forecasting}

Forecasting techniques are useful in terms of energy efficiency, because they
help to develop predictive control systems. This section introduces formal
aspects and forecasting modeling done for this work. Time series are data
series with trend and pattern repetition through time. They can be formalized as
a sequence of scalars from a variable $x$, obtained as the output of the observed process:

\vspace {-12pt}
\begin{equation}
\sequence{s}(x) = s_0(x), s_1(x), \ldots, s_{i-1}(x), s_{i}(x), s_{i+1}(x) \,
\end{equation}
a fragment beginning at position $i$ and ending at position $j$
will be denoted by $s_{i}^{j}(x)$.

Time series forecasting could be grouped as \emph{univariate forecasting} when
the system forecasts variable $x$ using only past values of $x$, and
\emph{multivariate forecasting} when the system forecasts variable $x$ using
past values of $x$ plus additional values of other variables. Multivariate
approaches could perform better than univariate when additional variables cause
variations on the predicted variable $x$, as is shown in the experimental section.

Forecasting models are estimated given different parameters: the number of past
values, the size of the future window, and the position in the future of the
prediction (future horizon). Depending on the size of the future window and how it
is produced~\cite{2012:JESA:taieb}, forecasting approaches are denoted as: \linebreak 
\emph{single-step-ahead forecasting} if the model forecasts only the next time
step; \emph{multi-step-ahead iterative forecasting} if the model forecasts only
the next time step, producing longer windows by an iterative process; and
\emph{multi-step-ahead direct forecasting}~\cite{Cheng_Tan_Gao_Scripps_2006} if
the model forecasts in one step a large future window of size $Z$. Following
this last approach, two different major model types exist:

\begin{itemize}
\item \emph{Pure direct}, which uses $Z$ forecasting models, one for each
 possible future horizon.
\vspace {-9pt}
\item \emph{Multiple input multiple output} (MIMO), which uses one model to
 compute the full $Z$ future window. This approach has several advantages due to the joint learning of inputs and outputs, which allows the model to learn
 the stochastic dependency between predicted values. Discriminative models, as
 \anns, profit greatly from this
 input/output mapping. Additionally, \anns are able to learn non-linear
 dependencies.
\end{itemize}

\subsection{Forecast Model Formalization}\label{sec:forecastmodel}

A forecast model could be formalized as a function $F$, which receives as inputs
the interest variable ($x_0$) with its past values until current time $t$ and a
number $C$ of covariates ($x_1,x_2,\ldots,x_C$), also with its
past values, until current time $t$ and produces a future window of size $Z$
for the given $x_0$ variable:

\vspace {-9pt}

\begin{equation}
\langle \hat{s}_{t+1}(x_0),\hat{s}_{t+2}(x_0),\ldots,\hat{s}_{t+Z}(x_0) \rangle =
F(\Omega(x_0), \Omega(x_1), \ldots, \Omega(x_C)) \, 
\end{equation}
$\Omega(x)=s_{t-I(x)+1}^t(x)$ being the $I(x)$ past values of
variable/covariate $x$.


The number of past values $I(x)$ is important to ensure good performance of the
model, however, it is not easy to estimate this number exactly. In this work, it
is proposed to estimate models for several values of $I(x)$ and use the model
that achieves better performance, denoted as \best. It is known in the machine learning community that ensemble methods achieve better
generalization~\cite{1991:neuralcomputing:jacobs,2006:icaisc:raudys,Yu20082623}.
Several possibilities could be found in the literature, such as vote combination,
linear combination (for which a special case is the uniform or mean combination),
or in a more complicated way, modular neural networks~\cite{1994:nn:Happel}.
Hence, it is also proposed to combine the outputs of all estimated models for each different
value of $I(x)$, following a linear combination scheme (the linear combination is also known as ensemble averaging),
which is a simple, but effective method of combination, greatly extended to the machine
learning community. Its major benefit is the reduction of overfitting problems
and therefore, it could achieve better performance than a unique \ann.
The quality of the combination depends on the correlation of the \anns,
theoretically, as the more decorrelated the models are, the better the
combination is. In this way, different input size
$I(x)$ \anns were combined, with the expectation that they will be less correlated between themselves than other kinds of combinations, as modifying hidden layer size or other hyper-parameters.

A linear combination of forecasts models, given a set $F_{\model{1}},F_{\model{2}},\dots,F_{\model{M}}$ of $M$ forecast models, \linebreak with the same future window size ($Z$), follows this equation:

\vspace {-6pt}

\begin{equation}
\langle \hat{s}_{t+1}(x_0),\hat{s}_{t+2}(x_0),\ldots,\hat{s}_{t+Z}(x_0) \rangle = \sum_{i=1}^M \combw_i
F_{\model{i}}(\Omega_i(x_0), \Omega_i(x_1), \ldots, \Omega_i(x_C)) \,
\end{equation}
where $\combw_i$ is the combination weight given to the model
$\model{i}$; and $\Omega_i(x)$ is its corresponding $\Omega$ function, as
described in Section~\ref{sec:forecastmodel}. The weights are constrained to sum
one, $\sum_{i=1}^M \combw_i = 1$. This formulation allows one to combine forecast
models with different input window sizes for each covariate, but all of them
using the same covariate inputs. Each weight $\combw_i$ will be estimated
following two approaches:

\begin{itemize}
\item Uniform linear combination: $\combw_i=\frac{1}{M}$ for $1 \leq i \leq
 M$. Models following this approach will be denoted as \combeq.

\vspace {-6pt}
\item Exponential linear combination (softmax):

 \begin{equation}
 \combw_i =
 \frac{exp(L^{-1}(\model{i}, \data))}{\sum_{i=1}^M exp(L^{-1}(\model{i},
 \data))}
 \label{eq:comb}
 \end{equation}
for $1 \leq i \leq M$, $L^{-1}(\model{i},\data) =
1/L(\model{i},\data)$ being an inverted loss-function (error function) value for the
model $\model{i}$, given the dataset $\data$. It will be computed using a
validation dataset. In this paper, the loss-function will be the mean absolute error
(MAE), defined in Section~\ref{sec:evalmeasures}, because it is more robust on
outlier errors than other quadratic error measures. This approach will be
denoted as~\comb.
\end{itemize}

\subsection{Evaluation Measures}\label{sec:evalmeasures}

The performance of forecasting methods over one time series could be assessed by
several different evaluation functions, which measure the empirical error of the
model. In this work, for a deep analysis of the results, three different error
functions are used: MAE, root mean square error (RMSE)
and symmetric mean absolute percentage of error (SMAPE). The error is computed
comparing target values for the time series $s_{t+1},s_{t+2},\ldots,s_{t+Z}$, and its corresponding time series prediction $\hat{s}_{t+1},\hat{s}_{t+2},\ldots,\hat{s}_{t+Z}$, using the model $\model{}$:

\vspace {-12pt}
\begin{eqnarray}
 \text{MAE}(\model{}, t) &=& \frac{1}{Z} \sum_{z=1}^{Z} \vert \hat{s}_{t+z}(x_0) -
 s_{t+z}(x_0) \vert\\
\vspace {6pt}
 \text{RMSE}(\model{}, t) &=& \sqrt{\frac{1}{Z} \sum_{z=1}^{Z} ( \hat{s}_{t+z}(x_0) -
 s_{t+z}(x_0) )^2}\\
\vspace {6pt}
 \text{SMAPE}(\model{}, t) &=& \frac{1}{Z} \sum_{z=1}^{Z} \frac{ \vert \hat{s}_{t+z} -
 s_{t+z} \vert}{(\vert \hat{s}_{t+z} \vert + \vert s_{t+z} \vert)/2} \times 100\, 
\end{eqnarray}

The results could be measured over all time series in a given dataset $\data$ as:
\vspace {-3pt}
\begin{equation}
 L^\star(\model{}, \data) = \frac{1}{\vert \data \vert} \sum_{t=1}^{\vert
 \data \vert} L(\model{}, t) \, 
\end{equation}
$\vert \data \vert$ being the size of the
dataset and $L = \{\text{MAE}, \text{RMSE}, \text{SMAPE}\}$, the loss-function defining MAE$^\star$, RMSE$^\star$, and SMAPE$^\star$.

\subsection{Forecasting Data Description}

One aim of this work is to compare different statistical methods to forecast
indoor temperature given previous indoor temperature values. The correlation
between different weather signals and indoor temperature will also be analyzed.

In our database, time series are measured with a sampling period of $T=1$
min. However, in order to compute better forecasting models,
each time series is sub-sampled with a period of $T^\prime=15$ min, computing the mean of
the last $T^\prime$ values (for each hour, this mean is computed at 0 min, 15
min, 30 min and 45 min). The output of this preprocessing is the
data series $s^\prime(x)$, where:

\vspace {-6pt}

\begin{equation}
s^\prime_i(x) =
\displaystyle{\frac{\displaystyle{\sum_{j=(i-1) T^\prime + 1}^{i T^\prime}
 s_j(x)}}{T^\prime}} \, 
\end{equation}

\vspace {6pt}

One time feature and five sensor signals were taken into consideration:

\begin{itemize}
\item Indoor temperature in degrees Celsius, denoted by variable $x=d$. This
 is the interesting \linebreak forecasted variable.
\vspace {-9pt}
\item Hour feature in Universal Time Coordinated (UTC), extracted from the
 time-stamp of each pattern, denoted by variable $x=h$. The hour of the day is
 important for estimating the Sun's position.
\vspace {-9pt}
\item Sun irradiance in $W/m^2$, denoted by variable
 $x=W$. It is correlated with temperature, because more irradiance will mean
 more heat.
\vspace {-9pt}
\item Indoor relative humidity percentage, denoted by variable $x=H$. The humidity
 modifies the inertia of the temperature.
\vspace {-9pt}
\item Indoor air quality in CO$_{2}$ ppm (parts per million), denoted by variable
 $x=Q$. The air quality is related to the number of persons in the house, and a
 higher number of persons means an increase in temperature.
\vspace {-9pt}
\item Raining Boolean status, denoted by variable $x=R$. The result of
 sub-sampling this variable is the proportion of minutes in sub-sampling period
 $T^\prime$, where raining sensor was activated with \verb+True+.
\end{itemize}

To evaluate the forecasting models' performance, three partitions of our dataset
were prepared: \linebreak a \emph{training partition} composed of $2017$ time series over $21$
days---the model parameters are estimated to reduce the error in this data;
a \emph{validation partition} composed of $672$ time series over seven days---this is
needed to avoid over-fitting during training, and also to compare and study the
models between themselves; training and validation were performed in March 2011; a \emph{test partition} composed of \linebreak $672$ time series over seven days in June 2011. At the end, the forecasting error in this partition will be provided,
evaluating the generalization ability of this methodology. The validation partition
is sequential with the training partition. The test partition is one week ahead of
the last validation point.

\section{Forecasting Methods}
\vspace {-12pt}
\subsection{Standard Statistical Methods}

Exponential smoothing and auto-regressive integrated moving average models
(ARIMA) are the two most widely-used methods for time series forecasting. These
methods provide complementary approaches to the time series forecasting
problems. Therefore, exponential smoothing models are based on a description of trend
and seasonality in the data, while ARIMA models aim to describe its
autocorrelations. Their results have been considered as a reference to compare to the
\ann results.

On the one hand, exponential smoothing methods are applied for forecasting. These
methods were originally classified by~\cite{1969:journal:pegels} according to
their taxonomy. This was later extended by~\cite{1985:journal:gardner}, \linebreak modified
by~\cite{2002:journal:hyndman} and extended by~\cite{2003:journal:taylor},
giving a total of fifteen methods. These methods could have different behavior
depending on their error component [\emph{A} (additive) and \textit{M} (multiplicative)],
trend component [\emph{N} (none), \textit{A} (additive), \textit{Ad} (additive damped),
 \textit{M} (multiplicative) and \textit{Md} (multiplicative damped)] and seasonal component
[\emph{N} (none), \textit{A} (additive) and \textit{M} (multiplicative)]. To select the best-fitting
models within this framework, each possible model was estimated for the training
partition, and the two best models were selected. To carry out this selection,
Akaike's Information Criterion (AIC) was used as suggested by some works in the
literature \cite{Billah2006239,snyder2009exponential}. The selected models were:
\linebreak the first model with multiplicative error, multiplicative damped trend and without
the seasonal component (MMdN model), and the second model with additive error, additive
damped trend and without the seasonal component (AAdN model). The MMdN model was
chosen for the validation partition in order to minimize the MSE.

On the other hand, ARIMA models were estimated. The widely known ARIMA approach was first introduced by Box and Jenkins \cite{Box.Jenkins1976} and provides
a comprehensive set of tools for univariate time series modeling and forecasting.
These models were estimated for our data with and without covariates.
\linebreak The last value of variable hour ($x=h$), codified as a factor---using 24 categories (0 to 23), ---and the hour as a continuous variable were used as covariates.

Either linear and quadratic form of this quantity were used, but linear performs
worst. Therefore, three model groups are used: ARIMA without covariates (ARIMA), with
covariate $x=h$ as a factor (ARIMAF) and with covariate $x=h$ as a quadratic
form (ARIMAQ). The best models for each group were estimated for the training
partition, and in all cases, the non-seasonal ARIMA(2,1,0) model was selected for
the ARIMA part of each model using AIC. The best results, in terms of MSE,
were obtained in models with covariate time as a factor and covariate time as a
quadratic form.

The forecast library in the statistical package R~\cite{Rcommander} was used for
these analyses.

\subsection{ANNs}

Estimation of \ann forecast models needs data preprocessing and normalization of
input/output values in order to ensure better performance results.

\subsubsection{Preprocessing of Time Series for ANNs}

The indoor temperature variable ($x=d$) is the interesting forecasted variable. In
order to increase model generalization, this variable is differentiated, and a
new $\sequence{s}^{\prime\prime}(x=d)$ signal sequence is obtained following
this equation:

\vspace {-9pt}
\begin{equation}
s^{\prime\prime}_i(x=d) = s^\prime_i(x) - s^\prime_{i-1}(x) \, 
\end{equation}

The differentiation of indoor temperature shows that is important to
achieve good generalization results, and it is based on previous work where undifferentiated data has been used~\cite{2012:kdir:zamora}.

The time series corresponding to sun irradiance ($x=W$), indoor relative
humidity ($x=H$), \linebreak air quality ($x=Q$) and rain ($x=R$) are normalized,
subtracting the mean and dividing by the standard deviation, computing new signal
sequences, $\sequence{s}^{\prime\prime}(x \in \{W,H,Q,R\})$:

\begin{equation}
s^{\prime\prime}_i(x \in \{W,H,Q,R\}) = \displaystyle{\frac{s^\prime_i(x) -
 \mean{\sequence{s}^\prime(x)}}{\sigma(\sequence{s}^\prime(x))}} \, 
\end{equation}
\vspace {-6 pt}

\noindent where $\mean{\sequence{s}^\prime(x)}$ is the mean value of the sequence;
$\sequence{s}^\prime(x)$ and $\sigma(\sequence{s}^\prime(x))$ is the standard
deviation. \linebreak These two parameters may be computed over the training dataset. For
the hour component ($x=h$), a different approach is followed. It is represented
as a locally-encoded category, which consists of using a vector with $24$
components, where $23$ components are set to 0, and the component that
indicates the hour value is set to 1. This kind of encoding avoids the big
jump between 23 and 0 at midnight, \linebreak but forces the model to learn the
relationship between adjacent hours. Other approaches for hour encoding could be
done in future work.

\subsubsection{ANN Description}

\anns has an impressive ability to learn complex mapping functions, as they are universal function approximators~\cite{bishop:95} and are widely used in
forecasting~\cite{Zhang199835,2006:energyandbuildings:ruano,Yu20082623,2011:energyandbuildings:escriva}.

\begin{figure}[H]
 \centering
 \includegraphics[width=0.5\columnwidth]{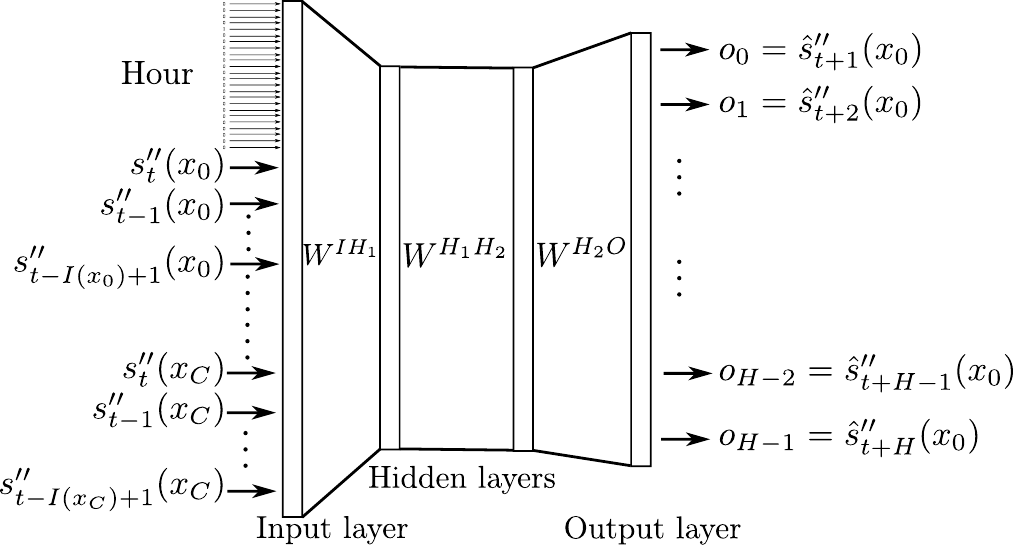}
 \caption{Artificial neural network (ANN) topology for time series forecasting.\label{fig:nn}}
\end{figure}

\anns are formed by one input layer, an output layer, and a few numbers of hidden
layers. Figure~\ref{fig:nn} is a schematic representation of an \ann with
two hidden layers for time series forecasting.
The inputs of the \ann are past values of covariates, and the output layer is formed
by the $Z$ future window predicted values, following the MIMO approach described
in Section~\ref{sec:forecasting}, which has obtained better accuracy in previous
experimentation~\cite{2012:kdir:zamora}.

The well-known error-backpropagation (BP)
algorithm~\cite{1988:nature:rumelhart} has been used in its on-line version to estimate the \ann weights, adding a momentum term and an L2
regularization term (weight decay).
Despite that theoretically algorithms more advanced than BP exists nowadays, BP is easier to implement at the empirical level, and a correct adjustment of
momentum and weight decay helps to avoid bad local minima.
The BP minimizes the mean
square error (MSE) function with the addition of the regularization term weight
decay, denoted by $\epsilon$, useful for avoiding over-fitting and improving
generalization:
\vspace {-3pt}
\begin{equation}
 E = \frac{1}{2Z} \sum_{i=1}^Z \left( \hat{s}_{t+i}(x_0) -
 s_{t+i}(x_0) \right)^2 + \frac{\epsilon}{2}\sum_{w_i \in \mathbf{\model}}
\displaystyle{w_i^2} \, 
\end{equation}
\vspace {6pt}
\noindent where $\mathbf{\model}$ is a set of all weights of
the \ann (without the bias); and $w_i$ is the value of the $i$-th weight.
\vspace {-6pt}

\section{Experimental Results}

Using the data acquired during the normal functioning of the house, experiments were
performed to obtain the best forecasting model for indoor
temperature. First, an exhaustive search of model hyper-parameters was done for each
covariate combination. Second, different models were trained for different
values of past size for indoor temperature $(x=d)$, and a comparison among
different covariate combinations and \ann \textit{vs}. standard statistical methods has
been performed. A comparison of a combination of forecasting models has also been performed. In all cases, the future window size $Z$ was set to $12$,
corresponding to a three-hour forecast.

A grid search exploration was done to set the best hyper-parameters of the system
and \ann topology, fixing covariates $x \in \{d,W,H,Q,R\}$ to a past size, $I(x)
= 5$ and $I(x=h)=1$, searching  \linebreak combinations of:
\vspace {-3 pt}

\begin{itemize}
\item different covariates of the model input;
\vspace {-9pt}
\item different values for \ann hidden layer sizes;
\vspace {-9pt}
\item learning rate, momentum term and weight decay values.
\end{itemize}
\vspace {-3 pt}

\begin{figure}[H]
 \begin{center}
 \includegraphics[width=0.7\textwidth]{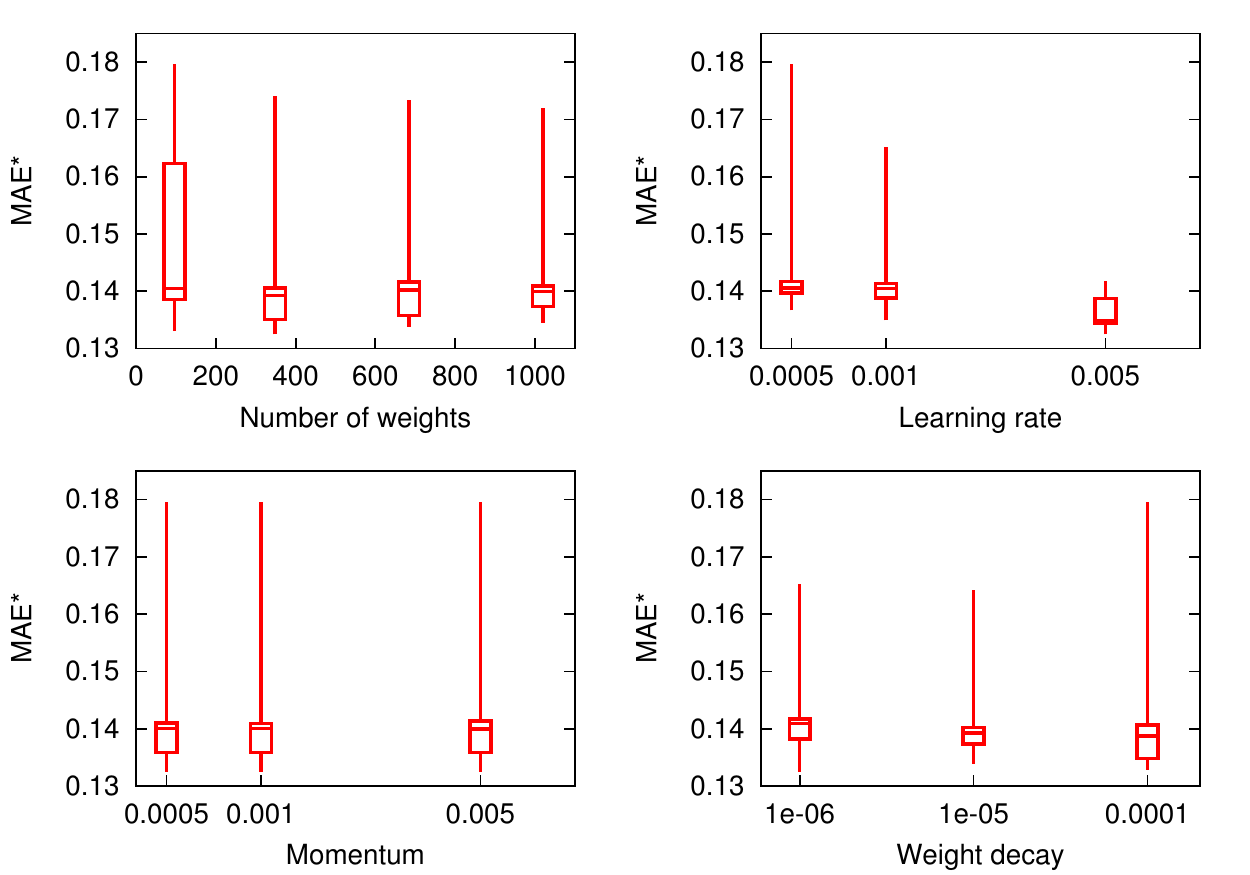}
 \end{center}
 \caption{Mean absolute error
(MAE)$^\star$ box-and-whisker plots for \anns with one hidden layer and
 the hyper-parameters of the grid search performed to optimize the \ann model,
 $d+h$.  The x-axis of the learning rate, momentum and weight decay are
 log-scaled.}
\label{fig:dhnumh1}
\end{figure}

\begin{figure}[H]
 \begin{center}
 \includegraphics[width=0.7\textwidth]{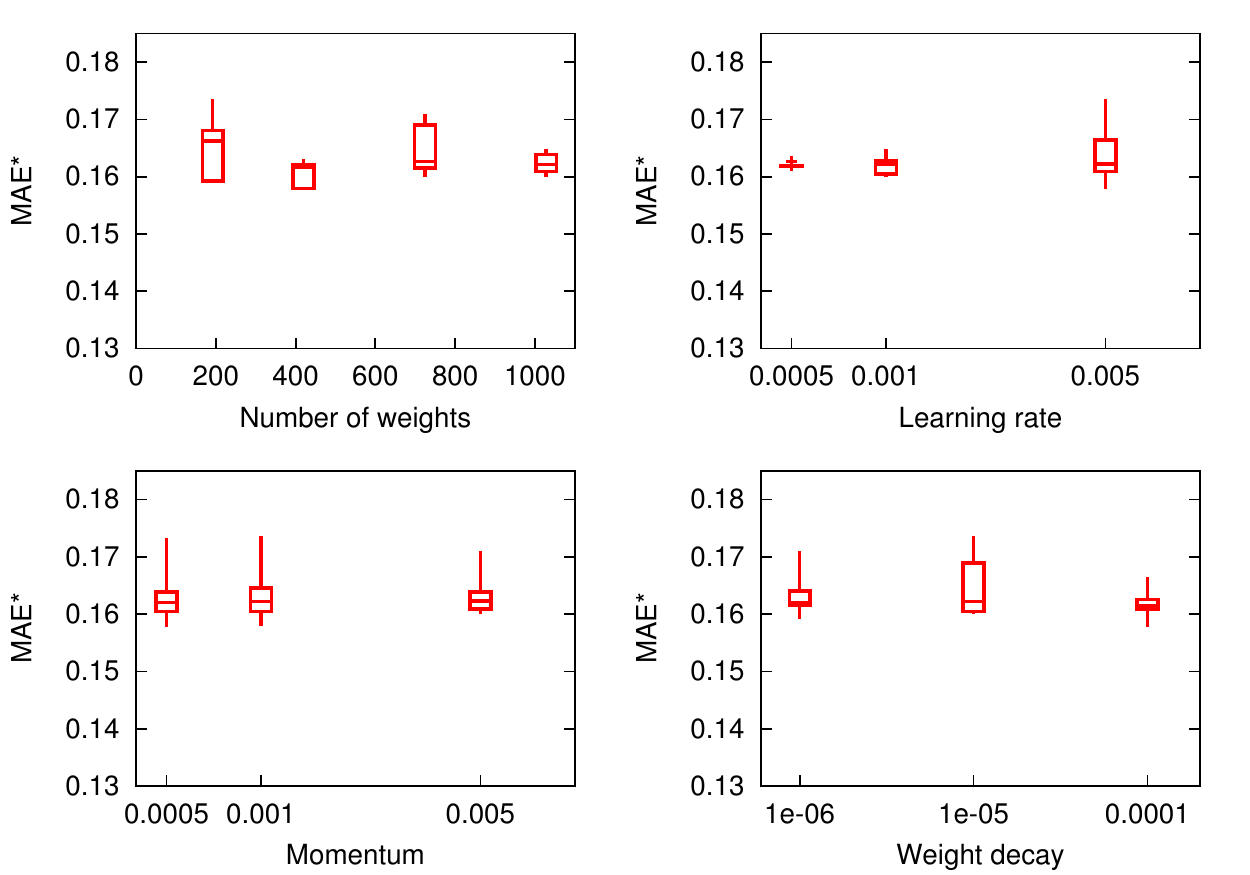}
 \end{center}
 \caption{MAE$^\star$ box-and-whisker plots for \anns with two hidden layers
 and the \protect\linebreak hyper-parameters of the grid search performed to optimize the \ann
 model, $d+h$. \protect\linebreak The x-axis of the learning rate, momentum and weight decay are
 log-scaled.}\label{fig:dhnumh2}
\end{figure}

Table~\ref{tab:modelparams} shows the best model parameters found by this grid
search. For illustrative purposes, Figures~\ref{fig:dhnumh1}~and~\ref{fig:dhnumh2} show box-and-whisker plots of the hyper-parameter grid
search performed to optimize the \ann model, $d+h$. They show big differences between one-
and two-hidden layer \anns, two-layered \anns being more difficult to train for this particular model. The learning rate shows a big impact in
performance, while momentum and weight decay seems to be less important. This
grid search was repeated for all the tested covariate combinations, and the
hyper-parameters that optimize MAE$^\star$ were selected in the rest of the
paper.

\begin{table}[H]
 \footnotesize \centering
 \begin{tabular}{cccccccc}
\toprule
 {\bf Covariates} & {\boldmath $\eta$} & {\boldmath $\mu$} & {\boldmath $\epsilon$} & {\bf Hidden layers}\\
 \midrule
 $d$   & $0.005\phantom{0}$ & $0.001$ & $1 \times 10^{-6}$ & $8$ tanh--$8$ tanh\\
 $d+W$   & $0.001\phantom{0}$ & $0.005$ & $1 \times 10^{-6}$ & $24$ tanh--$8$ tanh\\
 $d+h$   & $0.005\phantom{0}$ & $0.005$ & $1 \times 10^{-6}$ &{$8$ tanh}\\
 $d+h+W$   & $0.005\phantom{0}$ & $0.005$ & $1 \times 10^{-5}$ & $24$ tanh--$16$ tanh\\
 $d+h+H$   & $0.005\phantom{0}$ & $0.005$ & $1 \times 10^{-5}$ &{$16$ tanh}\\
 $d+h+R$   & $0.005\phantom{0}$ & $0.005$ & $1 \times 10^{-6}$ & $16$ logistic--$8$ logistic\\
 $d+h+Q$   & $0.0005$ & $0.005$ & $1 \times 10^{-4}$ &{$24$ logistic}\\
 $d+h+W+H$  & $0.005\phantom{0}$ & $0.005$ & $1 \times 10^{-5}$ &{$16$ tanh}\\
 $d+h+W+R$  & $0.005\phantom{0}$ & $0.005$ & $1 \times 10^{-6}$ & $16$ logistic--$8$ logistic\\
 $d+h+W+Q$  & $0.005\phantom{0}$ & $0.005$ & $1 \times 10^{-4}$ & $8$ tanh--$8$ tanh\\
 $d+h+W+Q+R$ & $0.005\phantom{0}$ & $0.005$ & $1 \times 10^{-4}$ & $24$ tanh--$8$ tanh\\
\bottomrule 
 \end{tabular}
 \normalsize
 \caption{Training parameters depending on the input covariates combination
 ($\eta$ is the learning rate, $\mu$ is the momentum term, and $\epsilon$ is weight
 decay).}\label{tab:modelparams}
\end{table}

\vspace {-18 pt}

\newpage

\subsection{Covariate Analysis and Comparison between Different Forecasting Strategies}

For each covariate combination, and using the best model parameters obtained
previously, different model comparison has been performed. Note
that the input past size of covariates is set to \linebreak $I(x \in \{W,H,Q,R\}) $= 5 time
steps, that is, $60$ min, and to $I(x=h)=1$. For forecasted variable $x=d$,
models with sizes $I(x=d) \in \{1,3,5,7,9,11,13,15,17,19,21\}$ were trained.

\begin{figure}[H]
 \centering
 \includegraphics[width=0.7\textwidth]{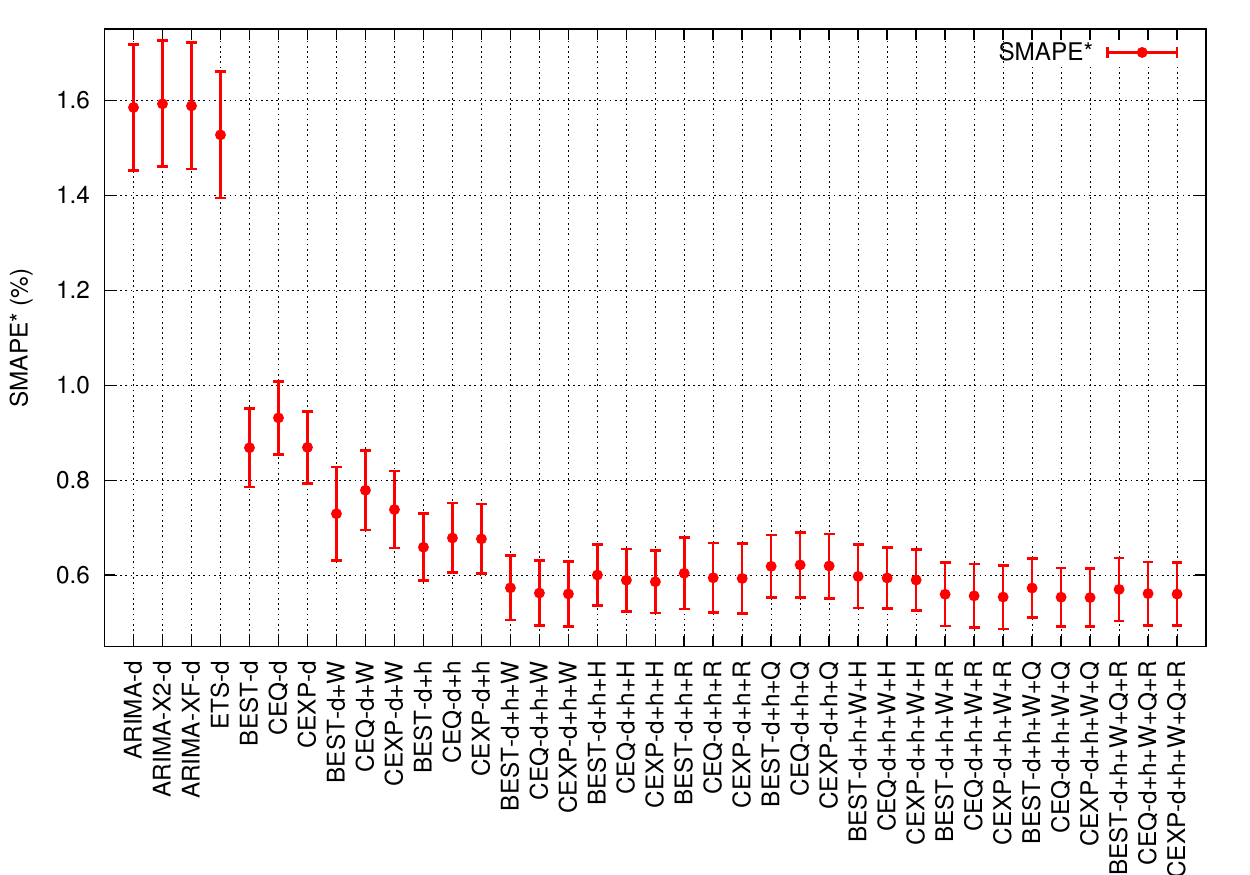}\\
 \caption{SMAPE$^\star$ error plot with $99\%$ confidence interval for models
 of Table~\ref{tab:valresults} on the validation partition.\label{fig:val-smape}}
\end{figure}

A comparison between \best, \combeq and \comb approaches was performed and
shown in Table~\ref{tab:valresults}. Figure~\ref{fig:val-smape} plots the same
results for a better confidence interval comparison. Table~\ref{tab:combs} shows
\combeq weights used in experimentation, obtained following
Equation~\ref{eq:comb} and using MAE$^\star$ as the \linebreak loss-function. From all these
results, the superiority of \anns \textit{vs}. standard statistical methods is clear,
with clear statistical significance and with a confidence greater than $99\%$. Different
covariate combinations for \ann models show that the indoor temperature
correlates well with the hour ($d+h$) and sun irradiance ($d+W$), and the combination
of these two covariates ($d+h+W$) improves the model in a significant way ($99\%$
confidence) with input $d+W$. The addition of more covariates is slightly
better in two cases ($d+h+W+R$ and $d+h+W+Q$), but the differences are not
important. With only the hour and sun irradiance, the \ann model has enough
information to perform good forecasting. Regarding the combination of models,
in some cases, the \comb approach obtains consistently better results than
\combeq and \best, but the differences are not important.

\begin{figure}[H]
 \centering
 \includegraphics[width=0.95\textwidth]{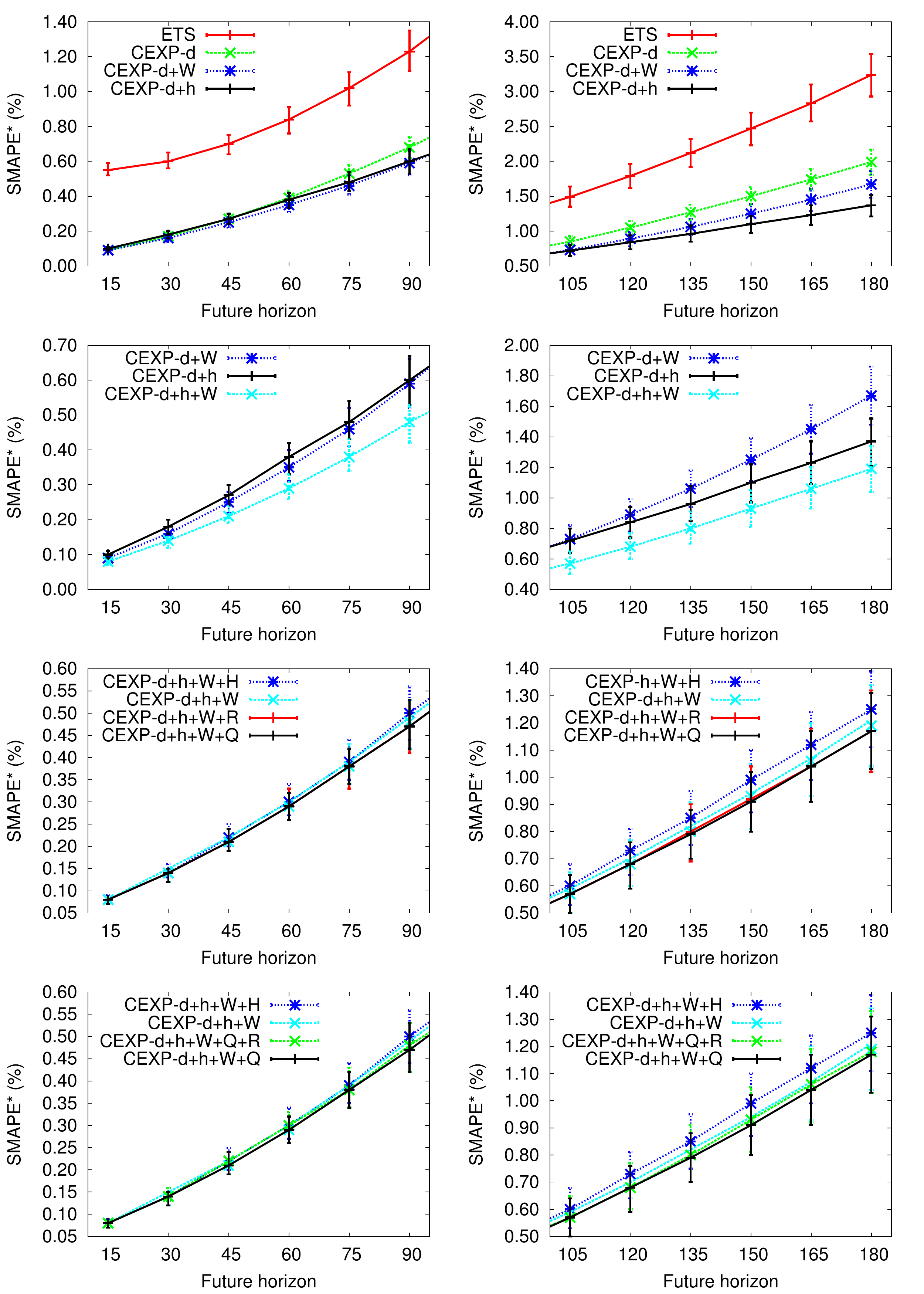}\\
 \caption{SMAPE$^\star$ error plot with $99\%$ confidence interval of each of
 the $Z=12$ future horizon predicted values (from 15 min forecast to 180
 min forecast.)\label{fig:val-180}}
\end{figure}

A deeper analysis could be done if comparing the SMAPE values for each possible
future horizon, as Figure~\ref{fig:val-180} shows. A clear trend exists: error
increases with the enlargement of the future horizon. Furthermore, an enlargement of
the confidence interval is observed with the enlargement of the future horizon. In all
cases, \ann models outperform statistical methods. For shorter horizons (less than or
equal to $90$ min), the differences between all \ann models are
insignificant. For longer horizons (greater than $90$ min), a combination of
covariates $d+h+W$ achieve a significant result (for a confidence of $99\%$)
compared with the $d+W$ combination. As was shown in these results, the addition
of covariates is useful when the future horizon increases, probably because the
impact of covariates into indoor temperature becomes stronger over time.

\begin{figure}[H]
 \centering
 \includegraphics[width=0.6\textwidth]{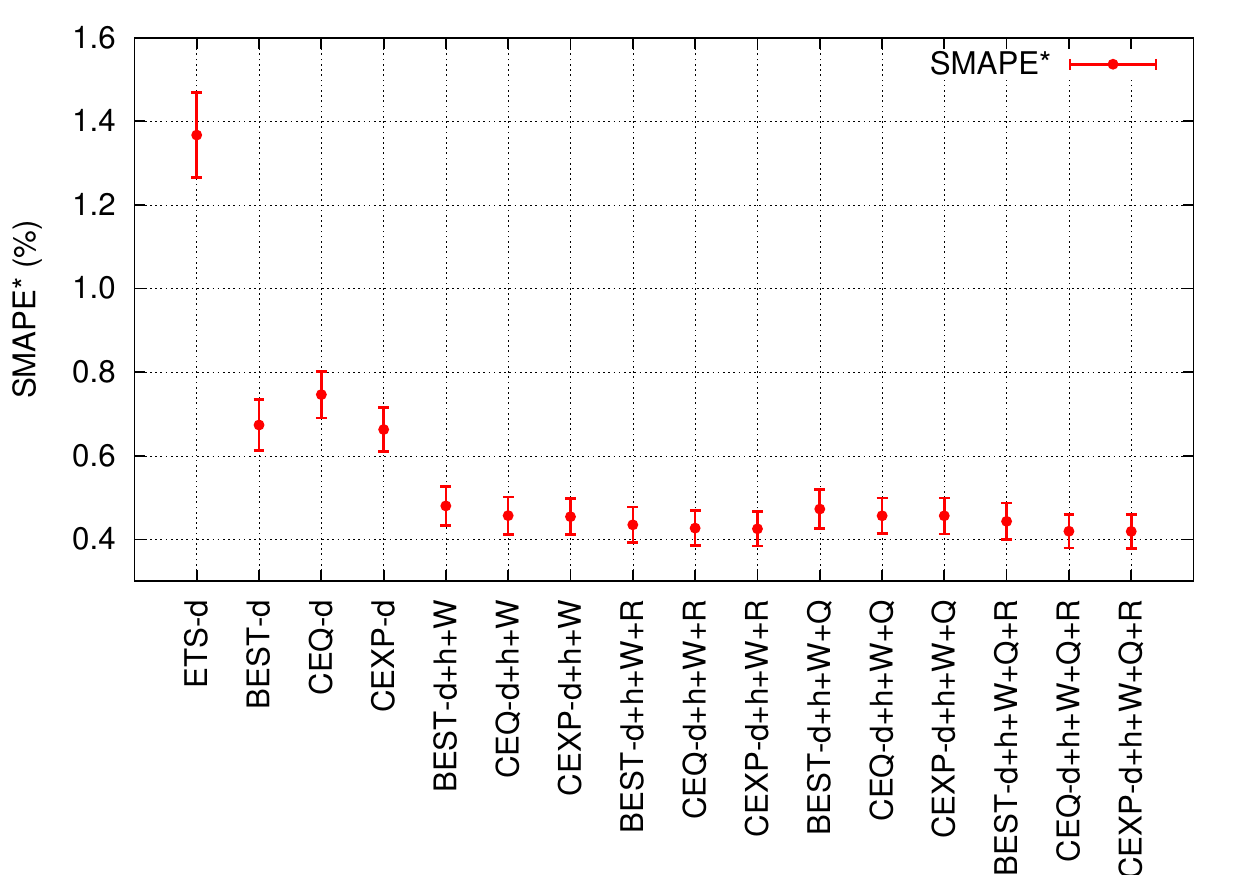}\\
 \caption{SMAPE$^\star$ error plot with the $99\%$ confidence interval for the models
 of Table~\ref{tab:testresults} in the test partition.\label{fig:test-smape}}
\end{figure}

Finally, to compare the generalization abilities of the proposed best models, the error
measures for the test partition are shown in Table~\ref{tab:testresults} and
Figure~\ref{fig:test-smape}. All error measures show better
performance in the test partition, even when this partition is two weeks ahead of
training and contains hotter days than the training and validation partitions. The reason for
this better performance might be that the test series has increasing/decreasing
temperature cycles that are more similar to the training partition than the cycles in the validation partition. The
differences between models are similar, and the most significant combination of
covariates is time hour and sun irradiance ($d+h+W$) following the \comb
strategy, achieving a SMAPE$^\star \approx 0.45\%$, MAE$^\star \approx 0.11$,
and RMSE$^\star \approx 0.13$.

\begin{table} [H]
\scriptsize \centering

 \begin{tabular}{cccccccccc}
 \toprule
 \bf {Model} & \multicolumn{3}{c}{\bf {SMAPE}$^\star (\%) {[lower, upper]}$} & \multicolumn{3}{c}{\bf {MAE}$^\star [lower, upper]$} & \multicolumn{3}{c}{\bf {RMSE}$^\star [lower, upper]$}\\
 \midrule
 \multicolumn{10}{c}{Standard statistical models}\\
 \hline
 ARIMA-$d$   & $1.5856$ & $[1.4528,$ & $1.7183]$ & $0.3099$ & $[0.2851,$ & $0.3348]$ & $0.3715$ & $[0.3413,$ & $0.4016]$\\
 ARIMAQ-$d+h^2$   & $1.5932$ & $[1.4607,$ & $1.7257]$ & $0.3113$ & $[0.2865,$ & $0.3362]$ & $0.3729$ & $[0.3428,$ & $0.4029]$\\
 ARIMAF-$d+h$   & $1.5888$ & $[1.4558,$ & $1.7219]$ & $0.3105$ & $[0.2857,$ & $0.3352]$ & $0.3721$ & $[0.3420,$ & $0.4022]$\\
 ETS-$d$   & $1.5277$ & $[1.3946,$ & $1.6607]$ & $0.3004$ & $[0.2753,$ & $0.3255]$ & $0.3648$ & $[0.3340,$ & $0.3957]$\\
 \hline
 \multicolumn{10}{c}{\ann models}\\
 \hline
 BEST-$d$   & $0.8687$ & $[0.7856,$ & $0.9517]$ & $0.1682$ & $[0.1524,$ & $0.1840]$ & $0.2109$ & $[0.1911,$ & $0.2306]$\\
 CEQ-$d$   & $0.9315$ & $[0.8545,$ & $1.0085]$ & $0.1802$ & $[0.1661,$ & $0.1944]$ & $0.2248$ & $[0.2072,$ & $0.2423]$\\
 CEXP-$d$   & $0.8695$ & $[0.7938,$ & $0.9452]$ & $0.1680$ & $[0.1541,$ & $0.1818]$ & $0.2109$ & $[0.1937,$ & $0.2280]$\\
 \hline
 BEST-$d+W$   & $0.7296$ & $[0.6311,$ & $0.8281]$ & $0.1418$ & $[0.1228,$ & $0.1608]$ & $0.1777$ & $[0.1544,$ & $0.2010]$\\
 CEQ-$d+W$   & $0.7792$ & $[0.6959,$ & $0.8625]$ & $0.1510$ & $[0.1353,$ & $0.1667]$ & $0.1888$ & $[0.1695,$ & $0.2082]$\\
 CEXP-$d+W$   & $0.7387$ & $[0.6576,$ & $0.8199]$ & $0.1430$ & $[0.1277,$ & $0.1582]$ & $0.1788$ & $[0.1601,$ & $0.1975]$\\
 \hline
 BEST-$d+h$   & $0.6593$ & $[0.5889,$ & $0.7298]$ & $0.1275$ & $[0.1143,$ & $0.1406]$ & $0.1549$ & $[0.1389,$ & $0.1708]$\\
 CEQ-$d+h$   & $0.6787$ & $[0.6055,$ & $0.7519]$ & $0.1312$ & $[0.1175,$ & $0.1449]$ & $0.1590$ & $[0.1425,$ & $0.1754]$\\
 CEXP-$d+h$   & $0.6768$ & $[0.6037,$ & $0.7498]$ & $0.1308$ & $[0.1172,$ & $0.1445]$ & $0.1586$ & $[0.1422,$ & $0.1750]$\\
 \hline
 BEST-$d+h+W$  & $0.5737$ & $[0.5058,$ & $0.6416]$ & $0.1121$ & $[0.0994,$ & $0.1248]$ & $0.1379$ & $[0.1222,$ & $0.1536]$\\
 CEQ-$d+h+W$   & $0.5625$ & $[0.4944,$ & $0.6306]$ & $0.1094$ & $[0.0966,$ & $0.1222]$ & $0.1348$ & $[0.1189,$ & $0.1506]$\\
 \rowcolor{lightgray}
 CEXP-$d+h+W$  & $0.5608$ & $[0.4927,$ & $0.6289]$ & $0.1091$ & $[0.0963,$ & $0.1218]$ & $0.1344$ & $[0.1187,$ & $0.1501]$\\
 \hline
 BEST-$d+h+H$  & $0.6006$ & $[0.5369,$ & $0.6642]$ & $0.1169$ & $[0.1050,$ & $0.1288]$ & $0.1429$ & $[0.1285,$ & $0.1573]$\\
 CEQ-$d+h+H$   & $0.5897$ & $[0.5240,$ & $0.6553]$ & $0.1142$ & $[0.1019,$ & $0.1264]$ & $0.1399$ & $[0.1250,$ & $0.1548]$\\
 CEXP-$d+h+H$  & $0.5864$ & $[0.5207,$ & $0.6521]$ & $0.1137$ & $[0.1014,$ & $0.1259]$ & $0.1393$ & $[0.1244,$ & $0.1543]$\\
 \hline
 BEST-$d+h+R$  & $0.6042$ & $[0.5292,$ & $0.6792]$ & $0.1170$ & $[0.1031,$ & $0.1309]$ & $0.1424$ & $[0.1255,$ & $0.1593]$\\
 CEQ-$d+h+R$   & $0.5947$ & $[0.5214,$ & $0.6680]$ & $0.1149$ & $[0.1014,$ & $0.1284]$ & $0.1410$ & $[0.1245,$ & $0.1575]$\\
 CEXP-$d+h+R$  & $0.5933$ & $[0.5196,$ & $0.6670]$ & $0.1146$ & $[0.1009,$ & $0.1282]$ & $0.1407$ & $[0.1241,$ & $0.1574]$\\
 \hline
 BEST-$d+h+Q$  & $0.6189$ & $[0.5526,$ & $0.6852]$ & $0.1200$ & $[0.1075,$ & $0.1325]$ & $0.1463$ & $[0.1311,$ & $0.1614]$\\
 CEQ-$d+h+Q$   & $0.6219$ & $[0.5539,$ & $0.6899]$ & $0.1208$ & $[0.1080,$ & $0.1336]$ & $0.1479$ & $[0.1324,$ & $0.1633]$\\
 CEXP-$d+h+Q$  & $0.6196$ & $[0.5518,$ & $0.6873]$ & $0.1203$ & $[0.1076,$ & $0.1331]$ & $0.1473$ & $[0.1319,$ & $0.1627]$\\
 \hline
 BEST-$d+h+W+H$  & $0.5977$ & $[0.5309,$ & $0.6646]$ & $0.1163$ & $[0.1037,$ & $0.1289]$ & $0.1434$ & $[0.1280,$ & $0.1588]$\\
 CEQ-$d+h+W+H$  & $0.5943$ & $[0.5304,$ & $0.6583]$ & $0.1155$ & $[0.1034,$ & $0.1275]$ & $0.1424$ & $[0.1276,$ & $0.1571]$\\
 CEXP-$d+h+W+H$  & $0.5899$ & $[0.5257,$ & $0.6540]$ & $0.1146$ & $[0.1025,$ & $0.1267]$ & $0.1413$ & $[0.1265,$ & $0.1561]$\\
 \hline
 BEST-$d+h+W+R$  & $0.5600$ & $[0.4935,$ & $0.6266]$ & $0.1090$ & $[0.0966,$ & $0.1214]$ & $0.1335$ & $[0.1183,$ & $0.1486]$\\
 CEQ-$d+h+W+R$  & $0.5568$ & $[0.4895,$ & $0.6240]$ & $0.1080$ & $[0.0955,$ & $0.1205]$ & $0.1328$ & $[0.1174,$ & $0.1482]$\\
 CEXP-$d+h+W+R$  & $0.5541$ & $[0.4872,$ & $0.6210]$ & $\mathbf{0.1076}$ & $[0.0951,$ & $0.1200]$ & $\mathbf{0.1323}$ & $[0.1169,$ & $0.1476]$\\
 \hline
 BEST-$d+h+W+Q$  & $0.5732$ & $[0.5111,$ & $0.6353]$ & $0.1118$ & $[0.1000,$ & $0.1236]$ & $0.1376$ & $[0.1231,$ & $0.1521]$\\
 CEQ-$d+h+W+Q$  & $0.5537$ & $[0.4921,$ & $0.6153]$ & $0.1079$ & $[0.0962,$ & $0.1196]$ & $0.1328$ & $[0.1184,$ & $0.1472]$\\
 CEXP-$d+h+W+Q$  & $\mathbf{0.5532}$ & $[0.4916,$ & $0.6148]$ & $0.1079$ & $[0.0962,$ & $0.1196]$ & $0.1328$ & $[0.1184,$ & $0.1472]$\\
 \hline
 BEST-$d+h+W+Q+R$ & $0.5704$ & $[0.5040,$ & $0.6369]$ & $0.1110$ & $[0.0984,$ & $0.1235]$ & $0.1363$ & $[0.1210,$ & $0.1517]$\\
 CEQ-$d+h+W+Q+R$ & $0.5615$ & $[0.4945,$ & $0.6285]$ & $0.1088$ & $[0.0964,$ & $0.1212]$ & $0.1340$ & $[0.1187,$ & $0.1492]$\\
 CEXP-$d+h+W+Q+R$ & $0.5606$ & $[0.4937,$ & $0.6275]$ & $0.1087$ & $[0.0963,$ & $0.1211]$ & $0.1337$ & $[0.1185,$ & $0.1490]$\\
 \bottomrule 
 \end{tabular}
 \normalsize
 \caption{Symmetric mean absolute percentage of error (SMAPE)$^\star$, MAE$^\star$ and root mean square error (RMSE)$^\star$ results on the validation
 partition comparing different models, input features and combination schemes
 with the $99\%$ confidence interval. BEST refers to the best past size \ann,
 CEQ refers to \combeq \anns, and CEXP refers to \comb \anns. Bolded face
 numbers are the best results, and the gray marked row is the most significant
 combination of covariates. ARIMA: auto-regressive integrated moving average models; ARIMAQ: ARIMA with covariate $x=h$ as a quadratic
form (ARIMAQ); ARIMAF: ARIMA with covariate $x=h$ as a factor.}
\label{tab:valresults}
\end{table}

\begin{table}[H]
\scriptsize \centering
 \begin{tabular}{cccccccccccc}
 \toprule
 &\multicolumn{11}{c}{{\bf \comb combination weights for every $d$ variable input size (min)}}\\ \cline{2-12}
\raisebox{2ex}[0pt]{{\bf Input covariates}} & $1 (15)$ & $3 (45)$ & $5 (75)$ & $7 (105)$ & $9 (135)$ & $11 (165)$ & $13 (195)$ & $15 (225)$ & $17 (255)$ & $19 (285)$ & $21 (315)$\\
 \midrule
 $d$   & $0.002$ & $0.044$ & $0.098$ & $\mathbf{0.142}$ & $0.092$ & $0.095$ & $0.082$ & $0.103$ & $0.100$ & $0.106$ & $0.135$ \\
\hline
 $d+W$   & $0.026$ & $0.020$ & $\mathbf{0.185}$ & $0.046$ & $0.069$ & $0.075$ & $0.104$ & $0.103$ & $0.124$ & $0.117$ & $0.131$ \\
\hline
 $d+h$   & $\mathbf{0.123}$ & $0.066$ & $0.099$ & $0.085$ & $0.092$ & $0.091$ & $0.091$ & $0.097$ & $0.084$ & $0.084$ & $0.088$ \\
\hline
 $d+h+W$   & $0.040$ & $0.112$ & $\mathbf{0.137}$ & $0.072$ & $0.078$ & $0.100$ & $0.107$ & $0.120$ & $0.083$ & $0.075$ & $0.075$ \\
\hline
 $d+h+H$   & $0.049$ & $0.058$ & $0.121$ & $\mathbf{0.127}$ & $0.095$ & $0.105$ & $0.114$ & $0.068$ & $0.100$ & $0.074$ & $0.090$ \\
\hline
 $d+h+R$   & $0.049$ & $0.052$ & $\mathbf{0.126}$ & $0.099$ & $0.078$ & $0.113$ & $0.104$ & $0.114$ & $0.102$ & $0.089$ & $0.076$ \\
\hline
 $d+h+Q$   & $0.084$ & $0.089$ & $0.105$ & $\mathbf{0.123}$ & $0.115$ & $0.103$ & $0.086$ & $0.077$ & $0.069$ & $0.073$ & $0.076$ \\
\hline
 $d+h+W+H$  & $0.062$ & $0.085$ & $0.071$ & $0.091$ & $0.123$ & $\mathbf{0.134}$ & $0.094$ & $0.082$ & $0.067$ & $0.121$ & $0.069$ \\
\hline
 $d+h+W+R$  & $0.048$ & $0.089$ & $\mathbf{0.142}$ & $0.078$ & $0.062$ & $0.116$ & $0.121$ & $0.092$ & $0.109$ & $0.087$ & $0.056$ \\
\hline
 $d+h+W+Q$  & $0.064$ & $0.101$ & $0.112$ & $0.097$ & $0.068$ & $0.088$ & $\mathbf{0.115}$ & $0.090$ & $0.085$ & $0.079$ & $0.101$ \\
\hline
 $d+h+W+Q+R$ & $0.042$ & $0.090$ & $\mathbf{0.136}$ & $0.098$ & $0.111$ & $0.089$ & $0.101$ & $0.072$ & $0.090$ & $0.085$ & $0.087$ \\
 \bottomrule 
 \end{tabular}
 \normalsize
 \caption{Combination weights of every input size of $d$ for the \comb models
 given tested covariates combinations. All co-variables have an input size of
 $5$ ($75$ min). \protect\linebreak Bold numbers are the best input sizes.}
\label{tab:combs}
\end{table}

\begin{table}[H]
 \scriptsize \centering

 \begin{tabular}{ccccccccccc}
 \toprule
 {\bf Model} & \multicolumn{3}{c}{{\bf SMAPE}$^\star (\%) [lower,upper]$} & \multicolumn{3}{c}{{\bf MAE}$^\star [lower,upper]$} & \multicolumn{3}{c}{{\bf RMSE}$^\star [lower,upper]$}\\
 \midrule
 ETS-d & $1.3669$ & $[1.2649,$ & $1.4688]$ & $0.3254$ & $[0.3023,$ & $0.3485]$ & $0.3930$ & $[0.3643,$ & $0.4218]$\\
 \hline
 BEST-$d$   & $0.6736$ & $[0.6128,$ & $0.7343]$ & $0.1604$ & $[0.1460,$ & $0.1748]$ & $0.2022$ & $[0.1844,$ & $0.2199]$\\
 CEQ-$d$   & $0.7462$ & $[0.6907,$ & $0.8016]$ & $0.1767$ & $[0.1638,$ & $0.1895]$ & $0.2203$ & $[0.2046,$ & $0.2360]$\\
 CEXP-$d$   & $0.6630$ & $[0.6101,$ & $0.7159]$ & $0.1572$ & $[0.1450,$ & $0.1694]$ & $0.1976$ & $[0.1824,$ & $0.2127]$\\
 \hline
 BEST-$d+h+W$  & $0.4802$ & $[0.4339,$ & $0.5266]$ & $0.1143$ & $[0.1035,$ & $0.1252]$ & $0.1382$ & $[0.1252,$ & $0.1512]$\\
 CEQ-$d+h+W$   & $0.4569$ & $[0.4127,$ & $0.5012]$ & $0.1090$ & $[0.0985,$ & $0.1195]$ & $0.1318$ & $[0.1193,$ & $0.1443]$\\
 \rowcolor{lightgray}
 CEXP-$d+h+W$  & $0.4546$ & $[0.4111,$ & $0.4982]$ & $0.1085$ & $[0.0981,$ & $0.1189]$ & $0.1312$ & $[0.1188,$ & $0.1437]$\\
 \hline
 BEST-$d+h+W+R$  & $0.4350$ & $[0.3925,$ & $0.4774]$ & $0.1034$ & $[0.0935,$ & $0.1132]$ & $0.1255$ & $[0.1136,$ & $0.1374]$\\
 CEQ-$d+h+W+R$  & $0.4271$ & $[0.3854,$ & $0.4688]$ & $0.1013$ & $[0.0916,$ & $0.1111]$ & $0.1225$ & $[0.1109,$ & $0.1341]$\\
 CEXP-$d+h+W+R$  & $0.4253$ & $[0.3837,$ & $0.4670]$ & $0.1010$ & $[0.0913,$ & $0.1108]$ & $0.1223$ & $[0.1107,$ & $0.1339]$\\
 \hline
 BEST-$d+h+W+Q$  & $0.4727$ & $[0.4258,$ & $0.5196]$ & $0.1127$ & $[0.1015,$ & $0.1238]$ & $0.1353$ & $[0.1223,$ & $0.1483]$\\
 CEQ-$d+h+W+Q$  & $0.4565$ & $[0.4136,$ & $0.4994]$ & $0.1092$ & $[0.0988,$ & $0.1195]$ & $0.1314$ & $[0.1192,$ & $0.1436]$\\
 CEXP-$d+h+W+Q$  & $0.4565$ & $[0.4134,$ & $0.4995]$ & $0.1091$ & $[0.0988,$ & $0.1195]$ & $0.1313$ & $[0.1192,$ & $0.1435]$\\
 \hline
 BEST-$d+h+W+Q+R$ & $0.4434$ & $[0.3997,$ & $0.4872]$ & $0.1051$ & $[0.0949,$ & $0.1153]$ & $0.1268$ & $[0.1147,$ & $0.1388]$\\
 CEQ-$d+h+W+Q+R$ & $0.4195$ & $[0.3792,$ & $0.4597]$ & $0.0996$ & $[0.0903,$ & $0.1090]$ & $0.1201$ & $[0.1090,$ & $0.1312]$\\
 CEXP-$d+h+W+Q+R$ & $\mathbf{0.4192}$ & $[0.3790,$ & $0.4595]$ & $\mathbf{0.0994}$ & $[0.0902,$ & $0.1087]$ & $\mathbf{0.1200}$ & $[0.1089,$ & $0.1311]$\\
 \bottomrule
 \end{tabular}

 \normalsize
 \caption{SMAPE$^\star$, MAE$^\star$ and RMSE$^\star$ results on test partition
 comparing the best models with the $99\%$ confidence interval. Bolded face
 numbers are the best results, and the gray marked row is the most significant
 combination of covariates.}\label{tab:testresults}
\end{table}

\vspace {-9 pt}

In order to perform a better evaluation, the conclusions above are compared with
mutual information (MI), shown in Table~\ref{tab:MI}. Probability densities have
been estimated with histograms, making the assumption of independence between
time points, which is not true for time series~\cite{2009:arxiv:papana}, but
is enough for our contrasting purpose. The behavior of the \anns is similar to the MI
study. Sun irradiance ($W$) covariates show high MI with indoor temperature
($d$), which is consistent with our results. Humidity ($H$) and air quality ($Q$)
MI with indoor temperature ($d$) is higher than sun irradiance, which seems
contradictory with our expectations. However, if we compute MI only during the day
(removing the night data points), the sun irradiance shows higher MI with indoor
temperature than other covariates. Regarding the hour covariate, it shows lower
MI than expected, probably due to the cyclical shape of the hour, which breaks abruptly
with the jump between 23 and 0, affecting the computation of histograms.

\begin{table}[H]
 \footnotesize \centering
 \begin{tabular}{cccccccc}
 \toprule
  {\bf Data}& {\bf Algorithm}& {\boldmath $d$} & {\boldmath $h$} & {\boldmath $W$} & {\boldmath $H$} & {\boldmath $R$} & {\boldmath $Q$}\\ 
\midrule
 & MI (for $d$) & $9.24$ & $4.44$ & $6.06$ & $8.95$ & $0.51$ & $7.70$\\
\cline {2-8}
 \raisebox{1.5ex}[0pt] {Validation set}&Normalized MI (for $d$) & $2.00$ & $1.48$ & $1.65$ & $1.95$ & $1.06$ & $1.82$\\
 \hline
 Validation set,  &MI (for $d$) & $8.23$ & $3.50$ & $8.11$ & $8.09$ & $0.58$ & $7.41$\\
\cline {2-8}
 removing night data points & Normalized MI (for $d$) & $2.00$ & $1.42$ & $1.98$ & $1.97$ & $1.07$ & $1.89$\\
 \bottomrule
 \end{tabular}
 \caption{Mutual Information (MI) and normalized MI between
 considered covariates and the indoor temperature, for the validation
 set.}\label{tab:MI}
\end{table}

\section{Conclusions}

An overview of the monitoring and sensing system developed for the \smlsystem
solar powered house has been described. This system was employed during the participation at the Solar Decathlon Europe 2012 competition. The research in
this paper has been focused on how to predict the indoor temperature of a house,
as this is directly related to HVAC system consumption. HVAC systems represent
$53.89\%$ of the overall power consumption of the \smlsystem house. Furthermore,
performing a preliminary exploration of the \smlsystem competition data, the energy
used to maintain temperature was found to be $30\%$--$38.9\%$ of the energy needed
to lower it. Therefore, an accurate forecasting of indoor temperature could
yield an energy-efficient control.

An analysis of time series forecasting methods for prediction of indoor
temperature has been performed. A multivariate approach was followed, showing
encouraging results by using \ann models. Several combinations of covariates,
forecasting model combinations, comparison with standard statistical methods
and a study of covariate MI has been performed. Significant improvements were
found by combining indoor temperature with the hour categorical variable and sun
irradiance, achieving a MAE$^\star \approx 0.11$ degrees Celsius (SMAPE$^\star
\approx 0.45\%$). The addition of more covariates different from hour and sun
irradiance slightly improves the results. The MI study shows that humidity and
air quality share important information with indoor temperature, but probably,
the addition of these covariates does not add different information from which is indicated by hour and sun irradiance. The combination of \ann models
following the softmax approach (\comb) produce consistently better forecasts,
but the differences are not important. The data available for this study was
restricted to one month and a week of a Southern Europe house. It might be interesting to
perform experiments using several months of data in other houses, as weather
conditions may vary among seasons and locations.

As future work, different techniques for the combination of forecasting models
could be performed. \linebreak A deeper MI study to understand the relationship
between covariates better would also be interesting. The use of second order methods to
train the \ann needs to be studied. In this work, for the \ann models, the hour
covariate is encoded using 24 neurons; other encoding methods will be studied, for example, using splines, sinusoidal functions or a neuron with
values between 0 and 23.

Following these results, it is intended to design a
predictive control based on the data acquired \linebreak from \anns, for example, from
this one that is devoted to calculating the indoor temperature, extrapolating this
methodology to other energy subsystems that can be found in a home.

\acknowledgements{Acknowledgments}

This work has been supported by Banco Santander and CEU Cardenal Herrera University 
through the project Santander-PRCEU-UCH07/12.

\conflictofinterests{Conflicts of Interest}

The authors declared not conflict of interest.

\bibliographystyle{mdpi}
\makeatletter
\renewcommand\@biblabel[1]{#1. }
\makeatother

\end{document}